\documentclass[default,iicol]{sn-jnl}% Default with double column layout
\usepackage{subcaption}

\theoremstyle{thmstyleone}%
\theoremstyle{thmstyletwo}%

\theoremstyle{thmstylethree}%

\begin{document}

\title[Graph Normalized-LMP Algorithm for Signal Estimation Under Impulsive Noise]{Graph Normalized-LMP Algorithm for Signal Estimation Under Impulsive Noise}


\author[1]{\fnm{Yi} \sur{Yan}}\email{y-yan20@mails.tsinghua.edu.cn}

\author[1]{\fnm{Radwa} \sur{Adel}}\email{jiayue20@mails.tsinghua.edu.cn}

\author*[1]{\fnm{Ercan Engin} \sur{Kuruoglu}}\email{kuruoglu@sz.tsinghua.edu.cn}

\affil*[1]{\orgdiv{Tsinghua-Berkeley Shenzhen Institute}, \orgname{Tsinghua University}, \orgaddress{ \city{Shenzhen}, \country{China}}}


\abstract{
In this paper, we introduce an adaptive graph normalized least mean pth power (GNLMP) algorithm for graph signal processing (GSP) that utilizes GSP techniques, including bandlimited filtering and node sampling, to estimate sampled graph signals under impulsive noise. 
Different from least-squares-based algorithms, such as the adaptive GSP Least Mean Squares (GLMS) algorithm and the normalized GLMS (GNLMS) algorithm, the GNLMP algorithm has the ability to reconstruct a graph signal that is corrupted by non-Gaussian noise with heavy-tailed characteristics. 
Compared to the recently introduced adaptive GSP least mean pth power (GLMP) algorithm, the GNLMP algorithm reduces the number of iterations to converge to a steady graph signal.
The convergence condition of the GNLMP algorithm is derived, and the ability of the GNLMP algorithm to process multidimensional time-varying graph signals with multiple features is demonstrated as well. 
Simulations show the performance of the GNLMP algorithm in estimating steady-state and time-varying graph signals is faster than GLMP and more robust in comparison to GLMS and GNLMS.}




\maketitle

\section{Introduction}\label{sec1}

The effectiveness of graphs in representing irregular data made graphs popular in the era of the advancements in information and computation technologies today \cite{b1,b2}. However, with the technical advancements, we now can acquire data at a rate that is faster than ever before, resulting in a scenario that we cannot process data at the rate we collect data \cite{b1}. 
Research in graph signal processing (GSP) has shown to be the solution to resolve the problem of processing irregular data by extending classical signal processing techniques such as Fourier transform and wavelet transform to graphs utilizing the spectral graph theory \cite{b1,b2,b3,b4,bib_Wavelets,bib_GFT}. 
GSP-inspired ideas have a broad area of applications in various fields of study such as analysing brain signals \cite{bib_brain}, monitoring 5G Networks \cite{b5}, modeling temperature data \cite{bib_NLMS}, making protien-protien interaction prediction \citep{bib_protein}, and modeling traffic events \cite{b7}. 
By defining graph convolution in neural networks using the Graph Fourier Transform (GFT), GSP has entered the field of deep learning; architectures such as the ChebNet and the graph convolutional network (GCN) are both deep learning architectures based on GSP backbones \cite{b9,b10}. 
In classical signal processing, adaptive filtering algorithms are capable of performing online estimation of a time-varying signal and tracking the signal under noise. 
One of the most popular adaptive algorithms is the Least Mean Squares (LMS) algorithm, along with algorithms that were derived based on the LMS such as the Normalized LMS (NLMS) algorithm and the Recursive Least Squares (RLS) algorithm, are recently introduced to the GSP field and applied on real data. 
This combination of adaptive algorithms and GSP techniques first demonstrated promising performance at online estimation of time-varying graph signals under noise corruption, resulting in the adaptive GSP least mean squares (GLMS) algorithm \cite{b5}. 
The adaptive GSP normalized LMS (GNLMS) algorithm \cite{bib_NLMS} and the adaptive GSP recursive least squares (GRLS) algorithm \cite{b11} are two algorithms that are extended based on the GLMS, and were proposed as improved versions with faster convergence speed than the GLMS algorithm, resulting in parallelism between their classical signal processing counterparts.

Impulsive noise processes generally possess large or even infinite variance and the distribution generally have heavy-tailed characteristics, causing the least-squares-based algorithm to diverge and result in unstable estimation \cite{b17}. 
However, the least-squares-based approaches assume the data and the noise to be Gaussian, which is an oversimplification of noise scenarios seen in real life \cite{bib_lp}. 
The GLMS and the GNLMS algorithms both suffer from unstable behavior under the presence of impulsive noise with diverging variance under impulsive noise due to the fact that they are least-squares-based algorithms.
Existing literature have shown that ambient noise can have impulsive character in scenarios such as underwater communications \cite{b14}, PLC communications \cite{b15}, radar signal processing \cite{b16}, and astrophysics \cite{herranz04}. 
Various distribution families were proposed in the past to model impulsive noise, including student-t, generalized Gaussian, and $\alpha$-stable distributions \cite{bib_lp}. 
The $\alpha$-stable distribution family, which is the generalization of the Gaussian distribution, has been successfully used to model various kinds of noise, stands out due to conforming to the generalized central limit theorem \cite{bib_lp, 1201785}.
Applying minimum dispersion (MD) criterion by $lp$-norm optimization instead of $l_2$-norm optimization seen in least-squares approaches was suggested in previous literature to avoid the problem of instability of least-squares \cite{b18}. 
This led to GLMP algorithm \cite{b19}, which utilizes the MD criterion and provides robust estimation of graph signal under the presence of $\alpha$-stable noise \cite{b19}. 
Although the GLMP algorithm is able to estimate the graph signal corrupted by $\alpha$-stable noise, it suffers from slow convergence speed similar to the GLMS algorithm. 

The previously mentioned GLMS, GRLS, GNLMS, and GNLMP algorithms only operate on graph signals with one single feature defined on the nodes of the graph, but data in the real world is often multidimensional and multi-featured. 
There are many real-life scenarios where the data is multidimensional and can potentially benefit if modeled using multidimensional graph signals. 
For example, in \cite{bib_health} multiple sensors are placed at multiple locations of the human body to monitor various vital signs and body motion, which could potentially be modeled using multi-feature GSP by treating the vitals or the motions as the features and constructing the graph topology using the location of the sensors. 
In \cite{bib_EEG}, a graph-theoretic model is given for multi-channel EEG signal, but the data from multi-channel are reduced to 1 dimension using synchronization likelihood. 
The GNLMS algorithm in \cite{bib_NLMS} is used to conduct online estimation of temperature recordings from weather stations; it would more beneficial if other data such as air quality, wind speed, precipitation, and humidity could be estimated simultaneously with the temperature.
In \citep{bib_GCN_air}, air pollutants recordings in weather stations, including CO, NO2, O3, PM10, PM2.5, and SO2, are being monitored using a fusion of the Attention mechanism and the GCN, but the inherent bulkiness of Neural Networks caused by their high complexity prohibits them to be applied to low-cost applications. 
Thus, there is a need for online processing of graph signals with multiple features defined on a single graph topology at a low cost.

In this paper, we propose a novel adaptive algorithm for GSP: the graph normalized least mean $p^{th}$ power (GNLMP) algorithm. 
The GNLMP algorithm is derived based on the MD criterion and with the concept of spectral-domain normalization to speed up the estimation process. 
Our proposed GNLMP algorithm does not experience the instability of the GLMS or GNLMS algorithms caused by the diverging variance of impulsive noise. 
Instead of a fixed step-size parameter seen in the GLMP algorithm, the GNLMP algorithm uses a time-varying convergence matrix $\mathbf{M\left[k\right]}$ that significantly reduces the number of iterations to converge to a stable value. 
We also propose an approximate version of the GNLMP to reduce computation complexity and reduce the run time; a switching strategy is provided to select one of the two versions.
We expanded the formulation of our GNLMP algorithm so that it could process multiple graph signals defined over the same graph simultaneously and overcome the limitation of the graph signal dimension. 
The proposed GNLMP algorithm is being tested on the ability to reconstruct a sampled graph signal under $\alpha$-stable noise using both synthetic and real data, with the graph signal being one of the following three settings: single feature steady-state, single feature time-varying, or multi-feature time-varying.

We present the background information on GSP in Section \ref{sec:BACKGROUND}. 
In Section \ref{sec:Alrorithm}, we present the GNLMP algorithm, along with the computational complexity analysis, and the steady-state convergence analysis. An extension of the GNLMP algorithm to graph signals with multiple features is also discussed in Section \ref{sec:Alrorithm}. 
Experimental studies can be found in Section \ref{sec_results}. Section \ref{sec_conclusion} summarizes and concludes the work.

\section{Background}
\label{sec:BACKGROUND}
\subsection{Graph Signal Processing Basics}\label{sec_GSP}
Let us first define a graph $\mathcal{G}=(\mathcal{V}, \mathcal{E})$, where $\mathcal{V}$ is the set of $N$ nodes, and $\mathcal{E}$ is the set of edges. 
For a weighted graph, the edge weight from node $v_i$ to node $v_j$ is the  $(i,j)^{th}$ entry of the graph adjacency matrix $\mathbf{A}$, and 0 if there is no connection between two nodes. 
For an unweighted graph the edge weights are 1 if there is an edge between node $v_i$ to node $v_j$, and 0 otherwise. 
The degree matrix $\mathbf{D}$ of an undirected and unweighted graph is a diagonal matrix with the $i^{th}$ diagonal entry being the number of edges $v_i$ has. 
The degree matrix of an undirected and weighted graph is a diagonal matrix with the $i^{th}$ diagonal entry being the summation of edges weights of node $v_i$.
The graph Laplacian matrix $\mathbf{L}$ of an undirected graph is defined as $\mathbf{L=D-A}$. 
A graph signal $\boldsymbol{x}$ is a graph with the function value defined on the nodes. 

The GFT is defined using the eigenvector decomposition of $\mathbf{L}$, $\mathbf{L=U\Lambda U^\mathit{T}}$, with $\mathbf{U}$ being the orthonormal eigenvectors of $\mathbf{L}$. 
The GFT of graph signal $\mathit{x}$ is used to transform $\boldsymbol{x}$ from spatial-domain to spectral-domain and is defined as $\boldsymbol{s}=\mathbf{U}^\mathit{T}\boldsymbol{x}$, which is the projection of $\boldsymbol{x}$ onto $\mathbf{U}$. 
The inverse graph Fourier transform (IGFT) $\mathbf{\boldsymbol{x}=U\boldsymbol{s}}$ transform $\boldsymbol{s}$ from spectral-domain to spatial domain. 
GSP algorithms can benefit from sparsity defined in both the spatial domain and the spectral domain. 
A bandlimited graph signal is sparse in the spectral-domain \cite{bib_NLMS,b8}. 
To get a bandlimited representation $\boldsymbol{x_0}$ of a graph signal $\boldsymbol{x}$, we apply a bandlimiting filter $\Sigma$ based on a frequency set $\mathcal{F}$, $\Sigma$ where is a diagonal matrix with idempotent and self-adjoint properties defined as $\mathbf{\Sigma_{ii}} = 1$ if $i\subseteq\mathcal{F}$ and 0 otherwise.
Then, the filter $\Sigma$ is appplyed using a graph convolution operation $\mathbf{\boldsymbol{x_0} = B\boldsymbol{x}}$, where $\mathbf{B = U\Sigma U^\mathit{T}}$. 
To simplify the notation, we define $\mathbf{U_\mathcal{F}=U\Sigma}$ then drop the columns with all zeros, we have $\mathbf{B = U_\mathcal{F}U_\mathcal{F}^\mathit{T}}$, $\mathbf{I} = \mathbf{U_\mathcal{F}^\mathit{T}U_\mathcal{F}}$, and $\mathbf{U}_\mathcal{F}^T\boldsymbol{x} = \text{support}(\mathbf{\Sigma U^\mathit{T} \boldsymbol{x}})$. 
The bandlimitedness of a graph signal $\boldsymbol{x_0}$ with frequencies $\mathcal{F}$ provides us the relationship $\mathbf{\boldsymbol{x_0} = B\mathbf{\boldsymbol{x_0}}}$\cite{b8}.
A graph signal represented using only a few sampled nodes is sparse in the spatial domain and can be obtained based on a sampling set $\mathcal{S}\subseteq\mathcal{V}$.
The sampling operation is done by an idempotent and self-adjoint diagonal sampling matrix $\mathbf{D}_\mathcal{S}$, with the diagonal entries equal to 1 when a node is sampled based on a sampling set $S\subseteq\mathcal{V}$ and 0 otherwise.

\subsection{The $\alpha$-stable Distribution}\label{sec_alpha}
To model impulsive noise in accordance with previous literature \cite{b16}, we use the symmetric $\alpha$-stable distribution (S$\alpha$S), which is a generalization of the Gaussian distribution.  
The S$\alpha$S is governed by three parameters: the characteristic exponent $\alpha$ that acts as tail-shape parameter, the dispersion $\gamma$ that acts as scale factor, and the location parameter $\mu_{\alpha}$. 
The S$\alpha$S distribution obeys the central limit theorem where linear combinations of independent S$\alpha$S random variables belong still to S$\alpha$S. 
The parameter $\alpha$ controls the impulsiveness of S$\alpha$S; $\mu_{\alpha}$ is the mean when $1<\alpha\leq2$ and the median when $1<\alpha$.
The parameter $\gamma$ controls the deviation around the mean or median. 
Unless when $\alpha$ = 2, the variance of S$\alpha$S diverges. 
The S$\alpha$S distributions have no analytic PDF except when $\alpha = 1$ and for $\alpha = 2$, which are the Cauchy distribution and the Gaussian distribution respectively. However, the characteristic function of S$\alpha$S could be expressed analytically as 
\begin{equation}
    \phi(t)=\exp\left\{j\mu_{\alpha} t-\gamma|t|^\alpha\right\}.\label{eq1}
\end{equation}
The $lp$-norm optimization used in our GNLMP algorithm utilizes the minimization of the dispersion, which is equivalent to minimizing the $p^{th}$-order moment when $1<\alpha<2$, or the fractional lower order moment (FLOM) \cite{b8}
\begin{equation}
    \begin{split}
    \text{FLOM}(p,\alpha,\gamma)=\mathbb{E}\mathbf{\lvert{X}\rvert}^p=C\left(p,\alpha\right)\gamma^{p/\alpha},\\
    \text{with } C\left(p,\alpha\right)=\frac{2^{p+1}\Gamma\left(\frac{p+1}{2}\right)\Gamma\left(-\frac{p}{\alpha}\right)}{\alpha\sqrt{\pi}\Gamma\left(-\frac{p}{2}\right)},\label{FLOM}
    \end{split}
\end{equation}
where $\mathbb{E}$ is the expectation operation.

\section{Adaptive GNLMP Algorithm for GSP}
\label{sec:Alrorithm}
\subsection{GLMS, GNLMS, and GLMP Analyses}
Following the convention of the previous adaptive GSP algorithms, we consider a bandlimited graph signal $\boldsymbol{x_0}\subseteq\mathbb{R}^N$, and its noisy observation with missing node values at iteration $k$ to be  expressed as a sampled noisy graph signal $\boldsymbol{y}\left[k\right] = \mathbf{D_\mathcal{S}}\left(\boldsymbol{x_0}+\boldsymbol{w}\left[k\right]\right)$. 
In this paper, the noise $\boldsymbol{w}\left[k\right]$ is modeled using S$\alpha$S with $\alpha\in\left(1,2\right)$.
The case where $\alpha<1$ is not considered because $\alpha<1$ indicates highly impulsive behavior and is rarely seen in reality \cite{b19}.

Least-squares-based algorithms are used extensively due to their simplicity of implementation. 
At iteration k, using the current step estimate $\hat{\boldsymbol{x}}\left[k\right]$, the cost function  $J\left(\hat{\boldsymbol{x}}\left[k\right]\right)$ for GLMS minimizes the meas-squared error of the estimation \cite{b5}:
\begin{equation}
\begin{split}
        J\left(\hat{\boldsymbol{x}}\left[k\right]\right)=\mathbb{E}f(\hat{\boldsymbol{x}}\left[k\right]), \\
        \text{where } f(\hat{\boldsymbol{x}}\left[k\right]) = \left\Vert\boldsymbol{y}\left[k\right]-\mathbf{D_\mathcal{S}B\hat{\boldsymbol{x}}\left[k\right]}\right\Vert_2^2.
        \label{lms_cost} 
\end{split}
\end{equation}
Using the cost function in \eqref{lms_cost} and the bandlimitedness property of $\boldsymbol{x}_0$ and $\hat{\boldsymbol{x}}\left[k\right]$, a convex optimization problem can be formed as shown below:
\begin{equation}
    \begin{split}
            \min_{\hat{\boldsymbol{x}}\left[k\right]} J(\hat{\boldsymbol{x}}\left[k\right])\\
        \textrm{s.t. } \mathbf{B}\hat{\boldsymbol{x}}\left[k\right] = \hat{\boldsymbol{x}}\left[k\right].
    \end{split}
    \label{minimization}
\end{equation}
The optimized solution of \eqref{minimization} could be obtained using stochastic gradient approaches. 
Knowing that $\mathbf{B}\hat{\boldsymbol{x}}\left[k\right] = \hat{\boldsymbol{x}}\left[k\right]$ for bandlimited graph signal, the spatial-domain update function of the GLMS algorithm is
\begin{equation}
    \begin{split}
    \hat{\boldsymbol{x}}\left[k+1\right]&=\hat{\boldsymbol{x}}\left[k\right]-\frac{\mu_{lms}}{2}\frac{\partial f(\hat{\boldsymbol{x}}\left[k\right])}{\partial \hat{\boldsymbol{x}}\left[k\right]}
    \\
    &=\hat{\boldsymbol{x}}\left[k\right]+\mu_{lms} \mathbf{BD_\mathcal{S}(\boldsymbol{y}\left[k\right]-\hat{\boldsymbol{x}}\left[k\right])}.
    \label{lms_ update}
    \end{split}
    % \hat{\boldsymbol{x}}\left[k+1\right]=\hat{\boldsymbol{x}}\left[k\right]+\mu_{lms} \mathbf{BD_\mathcal{S}(\boldsymbol{y}\left[k\right]-\hat{\boldsymbol{x}}\left[k\right])}.
    % \label{lms_ update}
\end{equation}
A step-size $\mu_{lms}$ is added to the GLMS to control the amount of the update of each iteration. 
Although simple to implement, the GLMS algorithm has two major drawbacks. 
First, the GLMS algorithm takes many iterations to convergence to a steady value \cite{bib_NLMS}. 
Second, the GLMS algorithm is derived with the assumption that noise follows the Gaussian distribution, but in reality, there are many non-Gaussian noise scenarios \cite{bib_lp,b14,b15,b16} that GLMS cannot handle.

In classical adaptive filtering, one possible solution to increase the convergence speed of the LMS algorithm is the normalization operation \cite{b20}. 
In GSP, the analogy of classical NLMS is the GNLMS algorithm;  instead of just having a fixed step-size, the GNLMS algorithm included a symmetric convergence matrix as normalization \cite{bib_NLMS}. 
The update function of the GNLMS algorithm could be expressed as
\begin{multline}
    \hat{\boldsymbol{x}}\left[k+1\right]=\\
    \hat{\boldsymbol{x}}\left[k\right]+\mu_{nlms} \mathbf{U_\mathcal{F}M_nU_\mathcal{F}^\mathit{T}D_\mathcal{S}(\boldsymbol{y}\left[k\right]-\hat{\boldsymbol{x}}\left[k\right])},\\
    \mbox{where }\mathbf{M}_\mathbf{n}=\left({\mathbf{U}_\mathcal{F}^T\mathbf{D}}_S\mathbf{U}_\mathcal{F}\right)^{-1},
    \label{NLMS_update}
\end{multline}
and $\mu_{nlms}$ is the step size parameter. 
Due to the inclusion of $\mathbf{M}$, $\mu_{nlms}$ and $\mu_{lms}$ will affect the update at each iteration differently. 
This is also the case for the GLMP algorithm and GNLMP algorithm that we will discuss in later parts of this paper.

The sensitivity to outliers of least squares due to its Gaussian noise assumption makes least-squares-based algorithms such as the GLMS and the GNLMS algorithms unstable under impulsive noise \cite{bib_lp}. 
In order to overcome this limitation, the GLMP algorithm was introduced in \cite{b19} as an improvement of the GLMS algorithm, modeling the noise using the S$\alpha$S. 
Instead of using least squares estimation, a $lp$-norm cost function is modified based on \eqref{lms_cost} to use the MD criterion to obtain a stable estimation under the presence of S$\alpha$S noise by setting $f(\hat{\boldsymbol{x}}\left[k\right]) = \left\Vert\boldsymbol{y}\left[k\right]-\mathbf{D_\mathcal{S}B\hat{\boldsymbol{x}}\left[k\right]}\right\Vert_p^p$. The parameter $p$ is chosen between $1<p<2$ to make the cost function differentiable \cite{b19}. 
The update function of the GLMP algorithm is derived using stochastic gradient approaches similar to \eqref{lms_ update}:
\begin{equation}
    \begin{split}
    \hat{\boldsymbol{x}}\left[k+1\right]
    &=\hat{\boldsymbol{x}}\left[k\right]+\mu_{lmp} \mathbf{BD}_\mathcal{S}\\
    &(\lvert\boldsymbol{y}\left[k\right]-\hat{\boldsymbol{x}}\left[k\right]\rvert^{p-1}\circ\text{Sign}(\boldsymbol{y}\left[k\right]-\hat{\boldsymbol{x}}\left[k\right])).
    \end{split} 
    \label{LMP_update} 
\end{equation}
The $\circ$ in \eqref{LMP_update} is the Hadarmad product between two matrices. The Sign() operation is 1 when the variable inside is less than 0, -1 when greater than 0, and 0 when exactly 0.

\subsection{GNLMP Algorithm Derivation}\label{sec_GNLMP_derivation}
Even though the GLMP algorithm is able to estimate the graph signal corrupted by S$\alpha$S, the GLMP algorithm still does not solve the problem of slow convergence speed compared to the GLMS algorithm. 
We propose the GNLMP algorithm that is based on the cost function of MD criterion and the idea of symmetric convergence matrix; the GNLMP algorithm can be formulated using stochastic gradients. 

We first transform \eqref{LMP_update} into the spectral-domain using GFT and replace $\mu_{lmp}$ with a time-varying convergence matrix $\mathbf{M}\left[k\right]$, leading to the spectral-domain update of the GNLMP algorithm:
\begin{multline}
    \hat{\boldsymbol{s}}\left[k+1\right]=\\
    \hat{\boldsymbol{s}}\left[k\right]+\mathbf{M}\left[k\right]\mathbf{U}_\mathcal{F}^T(\lvert\boldsymbol{e}\left[k\right]\rvert^{p-1}\circ\text{Sign}\left(\boldsymbol{e}\left[k\right]\right)),
    \label{GNLMP_update_d1}
\end{multline}
where $\boldsymbol{e}\left[k\right]=\mathbf{D}_S(\boldsymbol{y}\left[k\right]-\hat{\boldsymbol{x}}\left[k\right])$ is the current step spatial-domain estimation error.

In order to find $\mathbf{M}\left[k\right]$, following the derivations in \cite{bib_NLMS}, we define the \textit{a posteriori error} $\boldsymbol{\varepsilon}\left[k\right]=\mathbf{D}_S\left(\hat{\boldsymbol{x}}\left[k\right]-\mathbf{U}_\mathcal{F}\hat{\boldsymbol{s}}\left[k+1\right]\right)$ to measure the error between current step prediction and the next step prediction. We also define $\Delta{\widetilde{e}}^p=\left\Vert\boldsymbol{\varepsilon}\left[k\right]\right\Vert_p^p-\left\Vert\boldsymbol{e}\left[k\right]\right\Vert_p^p$ to be a measurement of close $\boldsymbol{\varepsilon}\left[k\right]$ is to $\boldsymbol{e}\left[k\right]$. 
The optimization problem in \eqref{minimization} can be seen as minimizing $\boldsymbol{e}\left[k\right]$, and in GNLMP  $\Delta{\widetilde{e}}^p$ is minimized as well. 
This minimization of $\Delta{\widetilde{e}}^p$ with respect to $\mathbf{M}\left[k\right]$ could be interpreted as using the spectral domain difference between $\hat{\boldsymbol{s}}\left[k+1\right]$ and $\hat{\boldsymbol{s}}\left[k\right]$ as spectral domain normalization of the update term \cite{bib_NLMS}, while keeping $\boldsymbol{e}\left[k\right]$ small based on \eqref{lms_cost}. Taking the derivative of $\Delta{\widetilde{e}}^p$ with respect to $\mathbf{M\left[k\right]}$, we have:
\begin{align}\label{GNLMP_derivation_1}
    &0=\mathbf{Q}^T
    \left(\mathbf{B}\lvert\boldsymbol{e}\left[k\right]\rvert^{p-1}\circ\text{Sign}(\boldsymbol{e}\left[k\right])\right)\nonumber
    \\&+\left(\lvert\boldsymbol{e}\left[k\right]\rvert^{p-1}\circ\text{Sign}(\boldsymbol{e}\left[k\right])\mathbf{B}\right)
   \mathbf{Q},\\\nonumber
   &\text{where }\mathbf{Q} = \\
   &\mathbf{D_\mathcal{S}U_\mathcal{F}M}[k]\mathbf{U}_\mathcal{F}^T\left(\lvert \boldsymbol{e}\left[k\right]\rvert^{p-1}\circ\text{Sign}(\boldsymbol{e}\left[k\right])\right)-\boldsymbol{e}[k].\nonumber
 \end{align}
Notice in \eqref{GNLMP_derivation_1} the two sides of the addition are transpose of each other, so when $\mathbf{Q}$ is zero, \eqref{GNLMP_derivation_1} is satisfied. 
Utilizing the property $\mathbf{I} = \mathbf{U_\mathcal{F}^\mathit{T}U_\mathcal{F}}$ and noticing that $\mathbf{D_s}$ is idempotent and self-adjoint, the expression for $\mathbf{M}[k]$ is 
\begin{equation}
    \mathbf{M}\left[k\right]=\left(\mathbf{U}_\mathcal{F}^T\mathbf{D}_S\text{diag}\left(\lvert\boldsymbol{y}\left[k\right]-\hat{\boldsymbol{x}}\left[k\right]\rvert^{p-2}\right)\mathbf{U}_\mathcal{F}\right)^{-1}.
    \label{M}
\end{equation}

Following the convention of classical adaptive filtering, we add a step size parameter $\mu$ to balance the convergence speed of the algorithm and the effectiveness of the update at each step. 
The spatial-domain update step of the adaptive GNLMP algorithm can be formalized as 
\begin{multline}
        \hat{\boldsymbol{x}}\left[k+1\right]=
        \hat{\boldsymbol{x}}\left[k\right]+\\
        \mu \mathbf{U_\mathcal{F}M}\left[k\right]\mathbf{U_\mathcal{F}^\mathit{T}}\left(\lvert\boldsymbol{e}\left[k\right]\rvert^{p-1}\circ\text{Sign}(\boldsymbol{e}\left[k\right])\right),
    \label{GNLMP_update}
\end{multline}
where $\mathbf{M}{k}$ is shown in \eqref{M}. 
Notice that this update function not only adaptively updates based on the error, but also has a time-varying parameter $\mathbf{M}\left[k\right]$, which is different from the GLMS, the GNLMS, and the GLMP algorithms as they only adaptively update the error. 
It is worth to mention that when we set $p=2$ in \eqref{GNLMP_update}, $\mathbf{M}\left[k\right]$ will reduce to $\mathbf{M}_n$, and  $\lvert\boldsymbol{e}\left[k\right]\rvert^{p-1}\circ\text{Sign}(\boldsymbol{e}\left[k\right])$ reduces to just $\boldsymbol{e}\left[k\right]$, which is exactly the update function of GNLMS in \eqref{NLMS_update}.
\subsection{Approximation of GNLMP}\label{sec:approximate}
Compared to the GLMP algorithm, the extra computations for calculating $\mathbf{M}\left[k\right]$ increases the run time of the GNLMP algorithm.
Since the problem is set up as convex optimization, it is safe to assume that after a few iterations the estimation error is mainly caused by the noise. 
Even though we do not have an analytical PDF for the S$\alpha$S, we can use the FLOM from \eqref{FLOM} to model the noise behavior.
Also, the sampling strategy we adopted from \cite{bib_NLMS} does not change as the algorithm progresses. 
Using these facts, we can approximate $\mathbf{M}\left[k\right]$ to be $\left(\mathbf{U}_\mathcal{F}^T\mathbf{D}_S\mathbf{RU}_\mathcal{F}\right)^{-1}$, with $\mathbf{R} = (E\lvert\boldsymbol{w}[k]\rvert^p)^{p-2}\mathbf{I}$ from \eqref{FLOM}.
 Using this approximation, we can combine all the time-independent terms of the GNLMP algorithm to form a matrix to perform the spectral domain filtering and normalization: $\mathbf{B}_{n} = \mathbf{U_\mathcal{F}}\left(\mathbf{U}_\mathcal{F}^T\mathbf{D}_S\mathbf{RU}_\mathcal{F}\right)^{-1}\mathbf{U_\mathcal{F}^\mathit{T}}$. Now the update function of GNLMP in \eqref{GNLMP_update} is approximated to be 
\begin{multline}
    \hat{\boldsymbol{x}}\left[k+1\right]
    =\hat{\boldsymbol{x}}\left[k\right]+\mu\mathbf{B}_{n}(\lvert\boldsymbol{e}\left[k\right]\rvert^{p-1}\circ\text{Sign}(\boldsymbol{e}\left[k\right])).
    \label{GNLMP_update_2}
\end{multline}

This formulation significantly reduces the number of operations done by our proposed GNLMP algorithm because now we can predefine $\mathbf{B}_n$ and calculate $\mathbf{B}_{n}$ only once. 
However, \eqref{GNLMP_update_2} loses the time-variability of $\mathbf{M}[k]$ that presents in \eqref{GNLMP_update}. 
To maintain the adaptiveness gained from the time-varying $\mathbf{M}\left[k\right]$ in \eqref{GNLMP_update} as well as maintaining the efficiency of \eqref{GNLMP_update_2}, we use a threshold-based switching between \eqref{GNLMP_update} and \eqref{GNLMP_update_2}. 
At step $k$, if the total amount of update magnitude at all sampled nodes is smaller than a certain threshold,  we switch from \eqref{GNLMP_update} to \eqref{GNLMP_update_2}, otherwise the update is \eqref{GNLMP_update}. 
Earlier in this section, we assumed that after a few iterations, $\boldsymbol{e}\left[k\right]$ is dominated by $\boldsymbol{w}\left[k\right]$, we set the the threshold to be $threshold = \lvert S\rvert\text{FLOM}(p-1,\alpha,\gamma)$, where $
\lvert\mathcal{S}\rvert$ is the cardinality of $\mathcal{S}$. 
The choice of threshold is not a strict requirement; it can be changed to other values to suit the need of the target application. The resulting GNLMP algorithm is shown in Algorithm \ref{algln2}.

\begin{algorithm}
\caption{Threshold based GNLMP}
\begin{algorithmic}[1]
\State $\mathbf{B}_{n} = \mathbf{U_\mathcal{F}}\left(\mathbf{U}_\mathcal{F}^T\mathbf{D}_S\mathbf{RU}_\mathcal{F}\right)^{-1}\mathbf{U_\mathcal{F}^\mathit{T}}$ \label{algln2}
\State $threshold=	\lvert\mathcal{S}\rvert\text{FLOM}(p-1,\alpha,\gamma)$
\While{within the iteration limit}
\If{\text{sum}$\left(\lvert\boldsymbol{e}\left[k\right]\rvert^{p-1}\right)<threshold$}
        \State update $ \hat{\boldsymbol{x}}\left[k+1\right]$ based on \eqref{GNLMP_update_2}
\Else
        \State update $ \hat{\boldsymbol{x}}\left[k+1\right]$ based on \eqref{GNLMP_update}
\EndIf
\EndWhile
    \label{algorithm}
\end{algorithmic}
\end{algorithm}

\subsection{Computational Complexity Analysis}
In this section, we will analyze the computational complexity of the proposed GNLMP algorithm. 
Comparing the approximated GNLMP update \eqref{GNLMP_update_2} with \eqref{LMP_update}, notice that both $\mathbf{B}_{n}$ and $\mathbf{B}$ have the same dimension $\mathbb{R}^{N\times N}$. 
$\mathbf{D}_\mathcal{S}(\lvert\boldsymbol{y}\left[k\right]-\hat{\boldsymbol{x}}\left[k\right]\rvert^{p-1}\circ\text{Sign}(\boldsymbol{y}\left[k\right]-\hat{\boldsymbol{x}}\left[k\right])) = \lvert\boldsymbol{e}\left[k\right]\rvert^{p-1}\circ\text{Sign}(\lvert\boldsymbol{e}\left[k\right]\rvert^{p-1})$, the difference of two expressions is only in the notation. 
As a result, the GNLMP approximation in \eqref{GNLMP_update_2} has the same computational complexity as the GLMP update in \eqref{LMP_update}. 
As for the update function in \eqref{GNLMP_update}, the extra computations comparing to \eqref{LMP_update} are the following: one diagonal matrix multiplication, one element-wise exponent, one matrix inverse, and three matrix multiplications. 
Matrix multiplication and matrix inverse are both $O(N^3)$ operations, and the element-wise exponent takes only $O(N\log(N))$. 
Due to the extra computations, the actual run time of \eqref{GNLMP_update} might be longer than \eqref{LMP_update}, but both the GLMP and the GNLMP are in fact dominated by the $O(N^3)$ operations. 

\subsection{Steady-state Convergence Behavior of GNLMP}
We would like to investigate the steady-state behavior of the GNLMP algorithm. For simplicity, we only analyze the approximated GNLMP update function in \eqref{GNLMP_update_2}. In reality, $\boldsymbol{y}\left[k\right]-\hat{\boldsymbol{x}}\left[k\right]$ is the error of the estimation, which is rarely zero. 
Then, for $\text{Sign}(\boldsymbol{y}\left[k\right]-\hat{\boldsymbol{x}}\left[k\right])\neq0$, \eqref{GNLMP_update_2} can be written as 
\begin{multline}
    \hat{\boldsymbol{x}}\left[k+1\right]
    =\hat{\boldsymbol{x}}\left[k\right]+
    \mu\mathbf{B}_{n}\mathbf{D}_\mathcal{S}\mathbf{R}_p(\boldsymbol{y}\left[k\right]-\hat{\boldsymbol{x}}\left[k\right]),
\end{multline}
where $\mathbf{R}_p = \lvert\text{diag}(\boldsymbol{y}\left[k\right]-\hat{\boldsymbol{x}}\left[k\right])\rvert^{p-2}$. Let the error update between $\boldsymbol{x}_0$ and $\hat{\boldsymbol{x}[k]}$ be $\tilde{\boldsymbol{x}}[k] = \hat{\boldsymbol{x}}[k] - \boldsymbol{x}_0$, then the error update function can be expressed as 
\begin{equation}
\begin{split}
    \tilde{\boldsymbol{x}}\left[k+1\right]
    =\tilde{\boldsymbol{x}}\left[k\right]+
    \mu\mathbf{B}_{n}\mathbf{D}_\mathcal{S}\mathbf{R}_p(\boldsymbol{w}\left[k\right]-\tilde{\boldsymbol{x}}\left[k\right])\\
    =\left(\mathbf{I}-\mu\mathbf{B}_n\mathbf{D}_\mathcal{S}\mathbf{R}_p\right)\tilde{\boldsymbol{x}}\left[k\right]+\mu\mathbf{B}_n\mathbf{D}_\mathcal{S}\mathbf{R}_p\boldsymbol{w}\left[k\right].
    \end{split}
    \label{error_update_2}
\end{equation}  
We can obtain the squared error of each update step based \eqref{error_update_2}, which leads to 
\begin{multline}
     \mathbb{E}\Vert\tilde{\boldsymbol{x}}\left[k+1\right]\Vert^2\\ =\mathbb{E}\Vert\tilde{\boldsymbol{x}}\left[k\right]\Vert^2_\mathbf{\Phi}+\mu^2\mathbb{E}\Vert\mathbf{B}_n\mathbf{D}_\mathcal{S}\mathbf{R}_p\boldsymbol{w}\left[k\right]\Vert^2.
\label{MSD_error}   
\end{multline}
In \eqref{MSD_error}, the notation $\Vert\tilde{\boldsymbol{x}}\left[k\right]\Vert^2_\mathbf{\Phi}$ refers to the weighted Euclidean norm $\tilde{\boldsymbol{x}}^T\left[k\right]\mathbf{\Phi}\tilde{\boldsymbol{x}}\left[k\right]$, and $\mathbf{\Phi} = \left(\mathbf{I}-\mu\mathbf{B}_n\mathbf{D}_\mathcal{S}\mathbf{R}_p\right)^T\left(\mathbf{I}-\mu\mathbf{B}_n\mathbf{D}_\mathcal{S}\mathbf{R}_p\right)$. 
Looking back into \eqref{error_update_2}, we could rewrite it into the following recursive relationship:
\begin{multline}
        \tilde{\boldsymbol{x}}\left[k+1\right] = \left(\mathbf{I}-\mu\mathbf{B}_n\mathbf{D}_\mathcal{S}\mathbf{R}_p\right)^k\tilde{\boldsymbol{x}}\left[0\right]
        \\+\sum_{i=0}^k\left(\mathbf{I}-\mu\mathbf{B}_n\mathbf{D}_\mathcal{S}\mathbf{R}_p\right)^k\boldsymbol{w}\left[i\right].
\end{multline}
As previously stated, we assume that the noise effect will dominate the error behavior and the noise is i.i.d. among the nodes, so $\mathbf{R}_p$ can be approximated to be $\lvert\boldsymbol{w}[k]\rvert^{p-2}$. We can now rewrite \eqref{MSD_error} in a recursive form:
\begin{align}
        &\mathbb{E}\Vert\tilde{\boldsymbol{x}}\left[k+1\right]\Vert^2\nonumber\\
        \label{MSD_error_2} &=\mathbb{E}\Vert\tilde{\boldsymbol{x}}\left[0\right]\Vert^2_\mathbf{\Phi^k}+\mu^2\sum_{i=0}^k\mathbb{E}\Vert\left(\mathbf{I}-\mu\mathbf{B}_n\mathbf{D}_\mathcal{S}\mathbf{R}_p\right)^k\Vert^2_\mathbf{G'}
        \\
        \label{MSD_error_3}
        &=\mathbb{E}\Vert\tilde{\boldsymbol{x}}\left[0\right]\Vert^2_\mathbf{\Phi^k}+\mu^2\sum_{i=0}^k\text{Tr}\left(\Phi^k\mathbf{G}\right)\\
        \label{MSD_error_4}
        &=\mathbb{E}\Vert\tilde{\boldsymbol{x}}\left[0\right]\Vert^2_\mathbf{\Phi^k}+\mu^2\sum_{i=0}^k\text{vec}\left(\mathbf{G}^T\right)\text{vec}\left(\Phi^k\right),
\end{align}
where $\mathbf{G} = \mathbf{B}_n\mathbf{D}_\mathcal{S}\mathbb{E}\lvert\boldsymbol{w}[k]\rvert^{2p-2}\mathbf{D}_\mathcal{S}\mathbf{B}_n$, and $\mathbf{G'} = \mathbf{B}_n\mathbf{D}_\mathcal{S}\lvert\boldsymbol{w}[k]\rvert^{2p-2}\mathbf{D}_\mathcal{S}\mathbf{B}_n$. 
Equation \eqref{MSD_error_3} is obtained from \eqref{MSD_error_2} by using the Trace property $\mathbb{E}\left\{\mathbf{X^TYX}\right\}=$Tr$\left(\mathbb{E}\left\{\mathbf{XX^TY}\right\}\right)$. Equation \eqref{MSD_error_4} is obtained by using the property Tr$(\mathbf{YX}) =$vec$(\mathbf{X^T})$vec$(\mathbf{Y})$, where vec$(\dot)$ is the operation of stacking each column of a matrix into a single colomn vector. 

In order for $\mathbb{E}\Vert\tilde{\boldsymbol{x}}\left[k+1\right]\Vert^2$ to converge in steady-state estimation of a graph signal, the condition $\Vert\left(\mathbf{I}-\mu\mathbf{B}_n\mathbf{D}_\mathcal{S}\mathbf{E}\{\mathbf{R}_p\right\})\Vert<1$ should be satisfied. 
Then in \eqref{MSD_error_4} the term $\mathbb{E}\Vert\tilde{\boldsymbol{x}}\left[0\right]\Vert^2_\mathbf{\Phi^k}$ will be 0 if $\tilde{\boldsymbol{x}}\left[0\right]$ is bounded, and the summation term $\mu^2\sum_{i=0}^k\text{vec}\left(\mathbf{G}^T\right)\text{vec}\left(\Phi^k\right)$ will be a converging geometric series. 
Given a symmetric matrix $\mathbf{Z}$, we have $\Vert\mathbf{Z}\Vert = \lvert\lambda_{max}\rvert$, where $\lambda_{max}$ is the largest eigenvalue of $\mathbf{Z}$. 
In the case of $\mathbf{Z} = \left(\mathbf{I}-\mu\mathbf{B}_n\mathbf{D}_\mathcal{S}\mathbf{E}\{\mathbf{R}_p\right\})$, we want $\Vert\mathbf{Z}\Vert<1$ for the GNLMP algorithm to converge. Since the only user defined variable is $\mu$, the following condition should be satisfied: 
\begin{equation}
    0 < \mu < \frac{2}{\lambda_{max}},
    \label{bound}
\end{equation}
where $\lambda_max$ is the maximum eigenvalue of $\mathbf{Z}$.

When condition \eqref{bound} is satisfied, the error or the deviation in mean-squared sense of a converging steady-state estimation can be obtained based on \eqref{MSD_error_4} using the property vec$\left(\mathbf{XYZ}\right) = \left(\mathbf{Z}^T \otimes \mathbf{X}\right)$vec$\left(\mathbf{Y}\right)$:
\begin{multline}
        \lim_{k \to \infty} \mathbb{E}\Vert\tilde{\boldsymbol{x}}\left[k\right]\Vert^2
        \\
        =\mu_{s}^2\text{vec}\left(\mathbf{G}\right)^T(\mathbf{I-Z^T\otimes Z})^{^-1}\text{vec}\left(\mathbf{I}\right),
    \label{MSD_simulation}
\end{multline}
The terms $\mathbb{E}\lvert\boldsymbol{w}[k]\rvert^{2p-2}$ in $\mathbf{G}$ and $\mathbb{E}\{\mathbf{R}_p\}$ in $\mathbf{Z}$ can be calculated using \eqref{FLOM}. 

\subsection{Processing Multi-feature Graph Signal Using GNLMP}
\begin{figure}[tb]
    \centering
    \begin{subfigure}{0.485\textwidth}
        \centering
        \includegraphics[width=\textwidth]{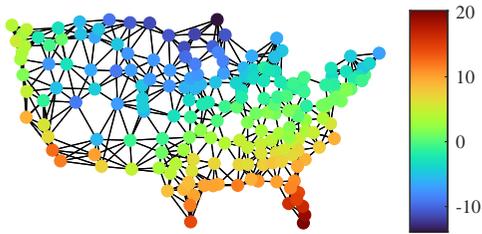}
        \caption{Temperature}
        \label{top_2}
    \end{subfigure}
    \begin{subfigure}{0.485\textwidth}
        \centering
        \includegraphics[width=\textwidth]{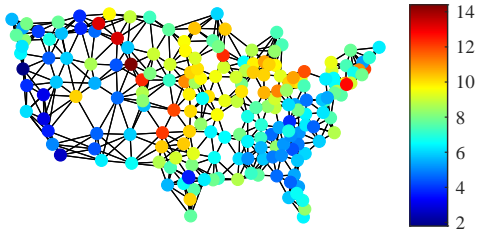}
        \caption{Wind speed}
        \label{top_2_winds}
    \end{subfigure}
\caption{The first time instance of two real time-varying graph signals defined on the same graph topology. }
\label{top_mult}
\end{figure}

Let us consider a bandlimited graph signal with $d$ features, where each feature is represented by a $N$ by 1 vector. 
The graph signal of interest $\mathbf{X}_0$ is a matrix of size $N$ by d, with each column being one feature. 
An illustration of such graph signal with $d=2$ is shown in Fig.~\ref{top_mult}, where we constructed a 7-nearest-neighbor graph of 197 weather stations using their geographic locations. There are two features in the graph signal, each representing the temperature in Fig.~\ref{top_2} and the wind speed Fig.~\ref{top_2_winds} recorded by the weather stations, and the graph signals are defined on the nodes. 
Such a multi-feature graph signal can be viewed as having two graph signals defined over the same graph topology. 
Estimation of such multi-feature graph signal using the adaptive GSP algorithms was not discussed in previous literature. 
One can try to treat different features as separate graph signals and process them independently.
For example, the GLMS algorithm will require each feature to be treated as a single one-dimensional graph signal, thus the estimation of $d$ features will be done using $d$ separate runs.  
We would like to expand the GNLMP algorithm to process such multidimensional graph signal $\mathbf{X}_0$ so all features are processed online simultaneously, instead of processing $d$ one dimensional $\boldsymbol{x}_0$s. 
For simplicity, we consider only the approximated GNLMP shown in \eqref{GNLMP_update_2} in this section. 
Let the noisy observation as iteration $k$ of $\mathbf{X_0}$ be $\mathbf{Y}\left[k\right]$, with the missing nodes modeled by sampling $\mathbf{D}_\mathcal{S}$, then we have $\mathbf{Y}\left[k\right] = \mathbf{D_\mathcal{S}}\left(\mathbf{X}_0+\mathbf{W}\left[k\right]\right)$, where $\mathbf{W}\left[k\right]$ is the noise i.i.d. among the nodes and features. We want to minimize the error of of the estimation $\hat{\mathbf{X}}[k]$ using the MD criterion, resulting in the cost function 
\begin{equation}
        J\left(\hat{\mathbf{X}}[k]\right)=\mathbb{E}\left\Vert\mathbf{Y}\left[k\right]-\mathbf{D_\mathcal{S}B_n\hat{\mathbf{X}}\left[k\right]}\right\Vert_p^p.
        \label{mult_cost} 
\end{equation}

The multi-feature GNLMP update function could be obtained by solving the following optimization problem:
\begin{equation}
    \begin{split}
            \min_{\hat{\boldsymbol{X}}\left[k\right]} J(\hat{\boldsymbol{X}}\left[k\right])\\
        \textrm{s.t. } \mathbf{B}_n\hat{\boldsymbol{X}}\left[k\right] = \hat{\boldsymbol{X}}\left[k\right].
    \end{split}
    \label{minimization2}
\end{equation}
Using the stochastic gradient approach similar to the single feature case, the expression for multi-feature GLMP update function is 
\begin{multline}
    \hat{\mathbf{X}}\left[k+1\right]
    =\hat{\mathbf{X}}\left[k\right]\\
    +\mathbf{B}_{n}(\lvert\mathbf{E}\left[k\right]\rvert^{p-1}\circ\text{Sign}(\mathbf{E}\left[k\right]))\mathbf{M}_\mu,
    \label{GNLMP_update_mult}
\end{multline}
where $\mathbf{E}\left[k\right] = \mathbf{D}_S(\mathbf{Y}\left[k\right]-\hat{\mathbf{X}}\left[k\right])$. The matrix $\mathbf{M}_\mu = $ diag${(\mu_1...\mu_d)}$ is added as step size matrix so each feature has a step size that can be tuned independently. 
Notice that when we process a multi-feature graph signal using the multi-feature GNLMP algorithm, the estimation of $d$ features is done simultaneously in an online fashion. We do not consider the $d$ features as separate graph signals in this setup, so we do not need to estimate them separately.  

\section{Experimental Results and Discussion}
\label{sec_results}
\subsection{Experiment Setup}
\label{sec_setup}
\begin{figure}[b]
\centerline{\includegraphics{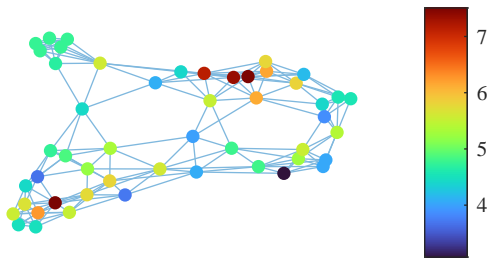}}
\caption{A random sensor graph and the graph signal.}
\label{top_1}
\end{figure}

The GNLMP algorithm shown in Algorithm \ref{algln2} is evaluated in Mean-squared deviation (MSD) performance under different experimental settings.
The instantaneous MSD at step $k$ is 
\begin{equation}
    MSD\left[{k}\right]=\mathbb{E} {\left\Vert\hat{\boldsymbol{x}}\left[k\right]-\boldsymbol{x_0}\right\Vert}_2^2.\label{MSD}
\end{equation}
In sections \ref{sec_mu_v}, \ref{sec_p_v}, and  \ref{sec_steady}, the experiments are conducted using a random sensor graph generated by Python PyGSP with N = 50 nodes. 
The generated instance used in this paper is shown in Fig.~\ref{top_1}. 
The frequency bands are $\lvert\mathcal{F}\rvert = 20$, selected using the techniques seen in \cite{bib_NLMS} to maximize spectral information, and the sampling strategy is the greedy strategy used in \cite{bib_NLMS} with $\lvert\mathcal{S}\rvert$ = 30. 

In section \ref{sec_tv}, a more realistic experiment is conducted by estimating a real time-varying graph signal of hourly weather collected from weather stations across the U.S. \cite{b24}. The graph signal is based on the graph topology in Fig.~\ref{top_2} with $N = 197$ and $t = 95$ time steps.
To compare the performance under time-varying setting, we replace the constant $\boldsymbol{x}_0$  with a time-varying graph signal $\boldsymbol{x}_0\left[k\right]$, then calculate the mean normalized MSD across time (NMSD) defined as
\begin{equation}
    % MSD_t\left[{k}\right]= \frac{1}{k}\sum_{n=1}^{k} MSD\left[{n}\right].
    NMSD_t\left[{k}\right]= \frac{1}{k}\sum_{n=1}^{k} MSD[k]/{\Vert\boldsymbol{x}_0[k]\Vert^2_2}.
\end{equation}
In section \ref{sec_mult}, a more challenging graph signal estimation is attempted. 
We would like to simultaneously estimate the two features of a real multidimensional time-varying graph signal; the $d = 2$ features are hourly temperature and average hourly wind speed collected from weather stations across the U.S. \cite{b24}. 
In Fig.~\ref{top_mult}, an illustration of  one  time  step  of  the  graph signal  is shown. 
For both the experiments on real data, we use the greedy sampling technique in \cite{bib_NLMS} with $\lvert\mathcal{S}\rvert=130$ to simulate missing node values and spatial domain sparsity. 
The frequency bands are selected using $\lvert\mathcal{F}\rvert = 125$ bands that have the maximum spectral information to provide spectral-domain sparsity \cite{bib_NLMS}.  

The baseline algorithms are the GLMS, the GNLMS, and the GLMP algorithms and will be selected based on the nature of the experiment.
In all experiments, $p=\alpha-0.05$ in the GNLMP algorithm and the GLMP algorithm as suggested in \cite{b17}. 
The results of all the experiments are averaged over 100 independent runs. 
The experiments were conducted on a computer with AMD Ryzen 5 3600 as CPU and RAM size of 32.0 GB using in MATLAB version 2020b.
\subsection{GNLMP Under Different Step Size}
\label{sec_mu_v}
\begin{figure}[bt]
    \centerline{\includegraphics[width=0.5\textwidth]{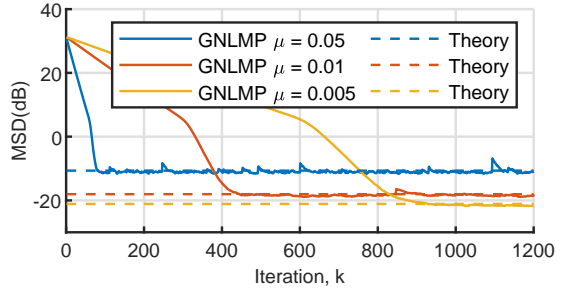}}
    \caption{MSD of steady-state graph signal estimation using GNLMP with different $\mu$ values.}
    \label{changing_mu}
\end{figure}

In this experiment, we want to verify that the algorithm works as intended. 
The graph signal shown in Fig~\ref{top_1} is corrupted by S$\alpha$S noise with $\alpha = 1.5$ and $\gamma = 0.1$. 
We run the GNLMP algorithm that is based on Algorithm \ref{algln2} to reconstruct the graph signal using step sizes $\mu = 0.05, 0.01,$ and $0.005$. 
The MSD of the reconstruction MSD is shown in Fig~\ref{changing_mu} along with the theoretical MSD calculated in \eqref{MSD_simulation}. 
From Fig~\ref{changing_mu}, we can conclude that as $\mu$ decreases, the algorithm will result in lower MSD, but needs more iterations to converge to a steady value. In other words, the convergence speed of GNLMP and the effectiveness of the update at each step can be tuned by tuning the step size $\mu$.
In Fig.~\ref{changing_mu} the theoretical MSD matches the actual MSD when the algorithm converges to a steady MSD.

\subsection{Graph Signal Estimation Under Various S$\alpha$S Noises}
\label{sec_p_v}
\begin{figure}[bt]
\centerline{\includegraphics[width=0.5\textwidth]{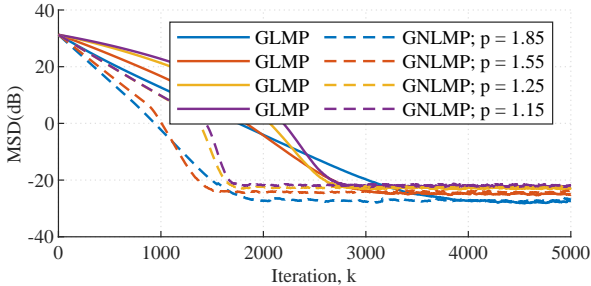}}
\caption{The MSD of estimating the graph signal under S$\alpha$S noises with $\alpha$ = 1.9, 1.5, 1.3, and 1.2. }
\label{p_v}
\end{figure}

The GNLMP algorithm is being compared with the GLMP algorithm in order to find out the effectiveness of the GNLMP algorithm at estimating a graph signal corrupted by various types of S$\alpha$S. 
We applied four different S$\alpha$S noises for $\boldsymbol{w}\left[k\right]$, with the parameters $\alpha = 1.9, 1.5, 1.3, 1.2$, and $\gamma = 0.1$. 
The step sizes of both algorithms the GNLMP algorithm and the GLMP algorithm are tuned so that both algorithms have approximately the same MSD values under each S$\alpha$S to eliminate the behavior differences between the step sizes. 
The number of iterations is set to 5000. 
The MSD of the estimation of the graph signal is shown in Fig.~\ref{p_v} and a table of run time for different settings is summarized in Table~\ref{tab1}. 
The steady convergence behavior of the MSD in Fig.~\ref{p_v} indicates the GNLMP algorithm is able to stably estimate the graph signal corrupted by various types of S$\alpha$S. 
Looking at the run time in Table~\ref{tab1}, we see that the GNLMP algorithm takes approximately the same amount of time as the GLMP algorithm to complete 5000 iterations. 
But to achieve the same MSD value the GLMP algorithm converges slower than the GNLMP algorithm, with $1.5\sim2$ times the iterations GNLMP algorithm needed to reach the same MSD as shown in Fig.~\ref{p_v}. 
\begin{table}[bt]
\caption{Run Time Comparison Between GLMP and GNLMP}
\begin{center}
\begin{tabular}{|c|c|c|c|c|}
\hline
 & $\alpha = 1.9$ & $\alpha = 1.6$ & $\alpha = 1.3$ & $\alpha = 1.2$\\
\hline
LMP & 0.0952(s) & 0.0950(s) & 0.0911(s) & 0.0909(s)\\
\hline
GNLMP & 0.1014(s) & 0.1060(s) & 0.1136(s) & 0.1183(s)\\
\hline
\end{tabular}
\label{tab1}
\end{center}
\end{table}

\subsection{Steady-state Performance of GNLMP Algorithm}
\label{sec_steady}
\begin{figure}[b]
\centerline{\includegraphics[width=0.5\textwidth]{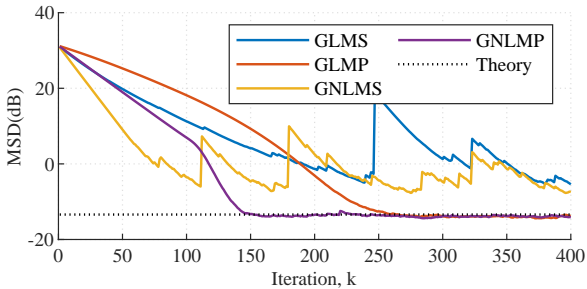}}
\caption{MSD of the steady-state graph signal estimation using the GLMS, GLMP, GNLMS, and GNLMP algorithm. }
\label{stationary_MSD}
\end{figure}

To measure  steady-state graph signal reconstruction quality of the GNLMP algorithm, the graph signal in Fig.~\ref{top_1} is corrupted by S$\alpha$S noise with parameters $\alpha = 1.5$ and $\gamma = 0.1$. 
The GNLMP algorithm is compared with the GLMS algorithm, the GLMP algorithm, and the GNLMS algorithm. 
Fig.~\ref{stationary_MSD} displays the MSD performance of estimating the graph signal along with the theoretical MSD of the GNLMP algorithm. 
In Fig.~\ref{stationary_MSD} both the GLMS and the GNLMS algorithms experience rapid MSD change and are unable to converge. 
This unstable behavior of the GLMS algorithm and the GNLMS is caused by the instability of using least-squares minimization under S$\alpha$S noise \cite{bib_lp}. 
The MD criterion used in the GNLMP algorithm does not suffer from the poor performance caused by the impulsiveness of S$\alpha$S noise. 
The GLMP algorithm and the GNLMP algorithm finish the 400 iterations in 0.0086 seconds and 0.0092 seconds respectively. 
Notice in Fig.~\ref{stationary_MSD}, the GNLMP algorithm reaches a steady MSD at iteration 150, whereas the GLMP algorithm reaches the same steady MSD at iteration 300, which confirmed that the GNLMP algorithm uses a fewer number of iterations to converge to a stable estimation. 

\subsection{Time-varying Graph Signal Estimation using GNLMP}
\label{sec_tv}
Extending the idea of steady-state graph signal reconstruction, we apply the GNLMP algorithm on a real time-varying graph signal of hourly temperature. 
There are two challenges in this experiment compared to the previous experiments. 
First, this experiment is now time-varying, which tests the ability of the GNLMP algorithm to conduct online estimation. 
Second, the data used in this experiment is gathered from the real world, which makes it not a perfectly bandlimited signal.
In Fig.~\ref{top_2}, an illustration of one time steps of the graph signal of hourly temperature is shown. The graph topology is generated using the approach as in \cite{bib_NLMS}, which could be summarized as treating each station as a node and connecting each station with its 7-nearest-neighbors. 
The edge connections are calculated using the geographical locations of the stations. 
Each weather station is represented as one node on the graph and the temperature recordings $\mathbf{x_0}[k]\in \mathbb{R}^{{N}\times{t}}$ is the graph signal of interest.
In this experiment, we only consider one single feature, the hourly temperature, out of the two features available so that we are able to compare our algorithm with the baseline algorithms. 

\begin{figure}[tb]
\centerline{\includegraphics[width=0.5\textwidth]{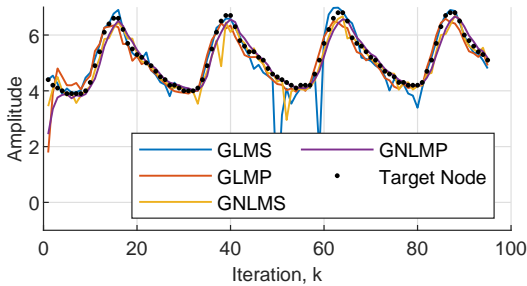}}
\caption{Estimation of a selected node from time-varying graph signal.}
\label{tv}
\end{figure}

\begin{figure}[tb]
\centerline{\includegraphics[width=0.5\textwidth]{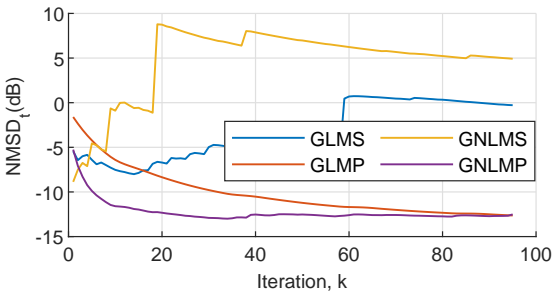}}
\caption{NMSD over time of time-varying estimation results of the entire graph.}
\label{tv_msd}
\end{figure}

To make the comparison between GLMP and GNLMP fair, both algorithms are tuned using grid search to a step size that has the best $MSD_t$ performance. 
In this experiment, the GNLMS algorithm is being compared to the GLMS algorithm, the GLMP algorithm, and the GNLMS algorithm. 
The S$\alpha$S noise is with parameters $\alpha = 1.5$ and $\gamma = 0.1$.
Fig.~\ref{tv} shows a reconstruction result of one selected node of the time-varying graph signal and Fig.~\ref{tv_msd} is the $NMSD_t$ of the experiment. 
Again, the reconstructions using the GLMS algorithm and the GNLMS algorithm are unstable due to diverging second-order moments under S$\alpha$S noise. 
This instability can be seen in both the signal estimation in Fig.~\ref{tv} and in the $NMSD_t$ in Fig.~\ref{tv_msd}. 
Compared to the GLMP algorithm, the GNLMP algorithm is able to reconstruct the time-varying graph signal with lower $NMSD_t$. 
\subsection{Multi-feature Time-varying Graph Signal Estimation Using GNLMP}
\label{sec_mult}
\begin{figure}[b]
\centerline{\includegraphics[width=0.5\textwidth]{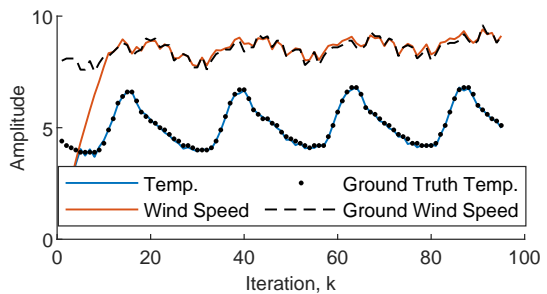}}
\caption{Estimation of the two features of a selected node from time-varying multi-feature graph signal.}
\label{mult}
\end{figure}

In this section, we further extend the time-varying graph signal estimation to multi-feature time-varying graph signal estimation. 
Fig.~\ref{top_mult}, shows the two features, the hourly temperature and hourly wind speed, of the multi-feature graph signal at the time point $k = 1$. 
Each weather station is represented as one node on the graph, resulting in a graph signal in the form $\mathbf{X_0}[k]\in \mathbb{R}^{d\times{N}\times{t}}$.
The graph signal  is being estimated using the GNLMP algorithm where the estimation results for both features are outputted simultaneously.
The experiment setup is essentially the same as Section \ref{sec_tv}, with $\lvert\mathcal{S}\rvert=130$ and $\lvert\mathcal{F}\rvert = 125$.
To take into account the fact that the two features are in different scales, two step sizes in $\mathbf{M}_\mu$ of \eqref{GNLMP_update_mult} are set to $\mu_1 = 0.55$ and $\mu_1 = 0.475$ respectively. 
The noise parameters of the S$\alpha$S noise is $\alpha = 1.5$ and $\gamma = 0.1$.
Fig.~\ref{mult} shows a reconstruction of both features at one selected node. 
By inspecting Fig.~\ref{mult}, we see that even though the two features each experience different magnitudes of change, the estimation for both features are accurate under S$\alpha$S noise. 
This indicates that the step size matrix $\mathbf{M}_u$ gives the GNLMP algorithm the freedom to update each feature with a different magnitude.
From Fig.~\ref{mult} we can see that the GNLMP algorithm is able to track the changes in both features in the multi-feature time-varying graph signal; the additional dimension of feature does not hinder the ability of the algorithm to make online predictions. 
\section{Conclusions}\label{sec_conclusion}
In this paper, we introduced and analyzed the adaptive GNLMP algorithm. To cope with the presence of S$\alpha$S noise, the GNLMP algorithm was derived based on the MD criterion. Compared to the traditional least-squares approaches, the proposed GNLMP algorithm does not suffer from the unstable estimation of least-squares caused by the heavy tail behavior of S$\alpha$S noise. The usage of a time-varying convergence matrix $\mathbf{M}\left[k\right]$ instead of a fixed step-size parameter makes it possible for the GNLMP algorithm to use fewer iterations to converge to a stable value than the recent GLMP algorithm. For steady-state estimations, the convergence condition for the GNLMP algorithm is provided. The GNLMP algorithm was also expanded to handle multi-feature graph signal instead of graph signals with only one feature. Experimenting with sampled and bandlimited graph signals corrupted with S$\alpha$S noise confirmed that the GNLMP algorithm is able to stably estimate the graph signal with faster convergence.
\section*{Statements and Declarations}
The authors have no competing interests to declare that are relevant to the content of this article.
\bmhead{Acknowledgments}
This work has been funded by High-end Foreign Expert Talent Introduction Plan under Grant G2021032021L.
\bmhead{Data Availability Statements}
The datasets generated during and/or analysed during the current study are available from the authors upon request.
\bibliographystyle{sn-mathphys}
\bibliography{sn-bibliography}
\vspace{25pt}

\end{document}